\documentclass[runningheads]{llncs}
\usepackage{graphicx}
\usepackage{float}
\usepackage{caption}
\begin{document}
\title{Ultra Low-Cost Two-Stage Multimodal System for Non-Normative Behavior Detection}
\titlerunning{Low-Cost Multimodal Non-Normative Behavior Detection}
%
\author{Albert Lu \and
Stephen Cranefield}
\authorrunning{A. Lu and S. Cranefield}
%
\institute{University of Cincinnati, USA \\
\email{lu2y4@mail.uc.edu} \\
\and
University of Otago, New Zealand \\
\email{stephen.cranefield@otago.ac.nz}
}

\maketitle              

\begin{abstract}
The online community has increasingly been inundated by a toxic wave of harmful comments. In response to this growing challenge, we introduce a two-stage ultra-low-cost multimodal harmful behavior detection method designed to identify harmful comments and images with high precision and recall rates. We first utilize the CLIP-ViT model to transform tweets and images into embeddings, effectively capturing the intricate interplay of semantic meaning and subtle contextual clues within both texts and images. Then in the second stage, the system feeds these embeddings into a conventional machine learning classifier like SVM or logistic regression, enabling the system to be trained rapidly and to perform inference at an ultra-low cost. By converting tweets into rich multimodal embeddings through the CLIP-ViT model and utilizing them to train conventional machine learning classifiers, our system is not only capable of detecting harmful textual information with near-perfect performance, achieving precision and recall rates above 99\% but also demonstrates the ability to zero-shot harmful images without additional training, thanks to its multimodal embedding input. This capability empowers our system to identify unseen harmful images without the need for extensive and costly image datasets. Additionally, our system quickly adapts to new harmful content; if a new harmful content pattern is identified, we can fine-tune the classifier with the corresponding tweets' embeddings to promptly update the system. This makes it well suited to addressing the ever-evolving nature of online harmfulness, providing online communities with a robust, generalizable, and cost-effective tool to safeguard their communities.

\end{abstract}

\section{Introduction}

With the increasing prevalence of online social communities, addressing harmful or non-normative behavior has become a critical concern. Harmful comments and images not only contribute to a toxic online environment but can also perpetuate harm and discrimination \cite{bib42,bib44}. Detecting and mitigating harmful behavior is crucial for fostering a safer and more inclusive online space. Previous studies on harmful behavior detection systems have not addressed the issue of multimodal representation learning for harmful behavior \cite{bib48,bib49}.

We introduce a two-staged multimodal harmful behavior detection method to combat harmful content online. This solution leverages advanced language models and established machine learning techniques to analyze both text and images, effectively detecting harmful content. It achieves this feat with minimal resource requirements, both for training and inference. Moreover, it has the ability to identify harmful images solely through a model trained on harmful text, eliminating the need for vast and costly image datasets.

Building upon the CLIP-ViT model \cite{bib1}, we generate multimodal embeddings for our collected harmful tweets, capturing both their semantic meaning and subtle contextual nuances. By further augmenting the dataset through rephrased tweets generated by the Mistral-7B-Instruct model \cite{bib2}, our system gains the ability to detect harmful content and the ability to detect it with greater precision. Subsequently, utilizing these rich representations, machine learning algorithms classify harmful comments and images with high accuracy, recall rate, and F1-score, as demonstrated in our experiments. These experimental results highlight the promise of our system in effectively addressing the pervasive challenge of harmful content within online communities.

Firstly, we began with generating tweet embeddings. We preprocessed 19,190 harmful tweets from the dataset created by \cite{bib4}. This data formed our training and testing set for harmful comments. For the non-harmful tweets, we obtained 10,000 normal and positive tweets from 80 topics using Twitter's API. Out of these, 6,252 tweets were in English. To expand our dataset, we utilized the Mistral-7B-Instruct model to generate an additional 10,825 rephrased non-harmful tweets. In total, our dataset contained 17,077 non-harmful tweets. All textual data was transformed into multimodal embeddings using the CLIP-ViT model. We performed dimensionality reduction with UMAP \cite{bib3} to facilitate visual analysis and exploration. The findings of this study demonstrate that integrating multimodal embeddings with traditional machine learning classifiers provides a cost-effective approach for identifying harmful content across text and images. This method is significantly more economical compared to alternative solutions that depend on fine-tuning resource-intensive models. Finally, we collected a dataset of regular and harmful images to assess the system's ability to generalize across modalities. Our experiments demonstrated that the harmful tweet embeddings and machine learning models trained on them cannot only identify textual harmful comments but also detect harmful images with zero-shot learning.

\section{Related Work}\label{sec2}

\textbf{Sentence Embedding.} Despite notable advancements, popular sentence embedding models like InferSent \cite{bib6} face limitations. Their treatment of each sentence as an isolated entity hinders their ability to capture crucial contextual information, ultimately impacting their ability to detect specific contexts. Universal Sentence Encoder \cite{bib7}, while offering significant progress, still suffers from limited contextual understanding and static embeddings. Although LASER \cite{bib9} and Sentence-BERT \cite{bib8} achieve state-of-the-art performance, they are both limited to textual data, rendering their embeddings inapplicable to multimodal data.

\vspace{\baselineskip}

\textbf{Dimensionality Reduction.} Autoencoders \cite{bib17}, including Variational Autoencoders (VAE) \cite{bib18}\cite{bib19} and Generative Adversarial Networks (GAN) \cite{bib20}, excel at learning interpretable latent representations and dimensionality reduction, but their computational complexity hinders their application in large-scale frameworks. In contrast, Principal Component Analysis (PCA) \cite{bib21} is computationally efficient and effective with data exhibiting linear relationships. The CLIP-ViT embeddings are known to be able to handle complex patterns and non-linear relationships, rendering PCA variants like Robust PCA \cite{bib25}, Kernel PCA \cite{bib26}, Sparse PCA \cite{bib27}, and Incremental PCA \cite{bib28} less suitable for capturing these intricate features.

\vspace{\baselineskip}

\textbf{Rephrased Comment Generation with Large Language Models.} Large language models have been widely used for generating rephrased comments. These models leverage their vast amount of memory and language understanding to generate alternative versions of given text \cite{bib22,bib23}. LLMs like GPT-4 \cite{bib10}, Palm 2 \cite{bib11}, and BARD \cite{bib12} can generate rephrased comments that maintain the original meaning while offering variations in wording and structure. However, their models are not open-sourced for research purposes.

Llama 2 \cite{bib13} on the other hand is open-sourced, cost-efficient, and has strong performance compared to previous open-source LLMs, but is not as efficient as the newer LLMs such as the Mistral model we are using in this paper. Falcon 180B \cite{bib14} is the current king of the jungle of LLMs with 180 billion parameters, trained on 3.5 trillion tokens. It requires a large amount of computing power for fine-tuning and inference, with 64 A100s needed for full fine-tuning, and around 16 for LoRa fine-tuning.

\vspace{\baselineskip}

\textbf{Multimodal Large Language Models (MM-LLMs)} MM-LLMs aim to learn joint representations from multiple modalities (text, image, audio and video) and have gained significant attention \cite{bib30,bib31,bib32,bib33} in recent years. 

The unprecedented ability of GPT-4V \cite{bib35} in processing arbitrarily interleaved multimodal inputs and the genericity of its capabilities together make GPT-4V a powerful multimodal generalist system. Nevertheless, it is not open-sourced for research purposes, so we did not use it in our work. The recently released open-source MM-LLM LLaVA \cite{bib36}, LLaVA-v1.5 \cite{bib37}, and Fuyu-8B \cite{bib38} are fast and incredibly powerful. However, their primary training objective is to function as a digital assistant that can answer questions based on the user's prompt and understand the images provided by the user, whereas our work focuses on multimodal embeddings that could precisely represent both textual and visual information.

\vspace{\baselineskip}

\textbf{harmful Comments Dataset From Online Social Community.} The Ruddit dataset \cite{bib39} contains comments from Reddit associated with fine-grained, real-valued scores ranging from -1 (totally normal comment) to 1 (indicating maximum harmfulness). However, it only provided post IDs and not the text of a post and while Reddit’s API allows retrieval of a comments given its post ID, we were only able to extract a few of them that still exist on Reddit, as most of the harmful comments have been deleted by the platform. 

We tried gathering normal comments through Reddit's API, focusing on those with a positive score of at least 3 (indicating community-perceived positivity), but we encountered limitations. Reddit's 1,000 comment crawl limit is insufficient for constructing an adequate amount of training data. Additionally, maintaining consistency in social norms would be optimal by using comments from a single community. Therefore, collecting both normal and harmful tweets from Twitter emerged as a better solution.

\section{Multimodal harmful Behavior Detection System}\label{sec3}

\begin{figure}[H]
  \centering
  \includegraphics[width=0.98\textwidth]{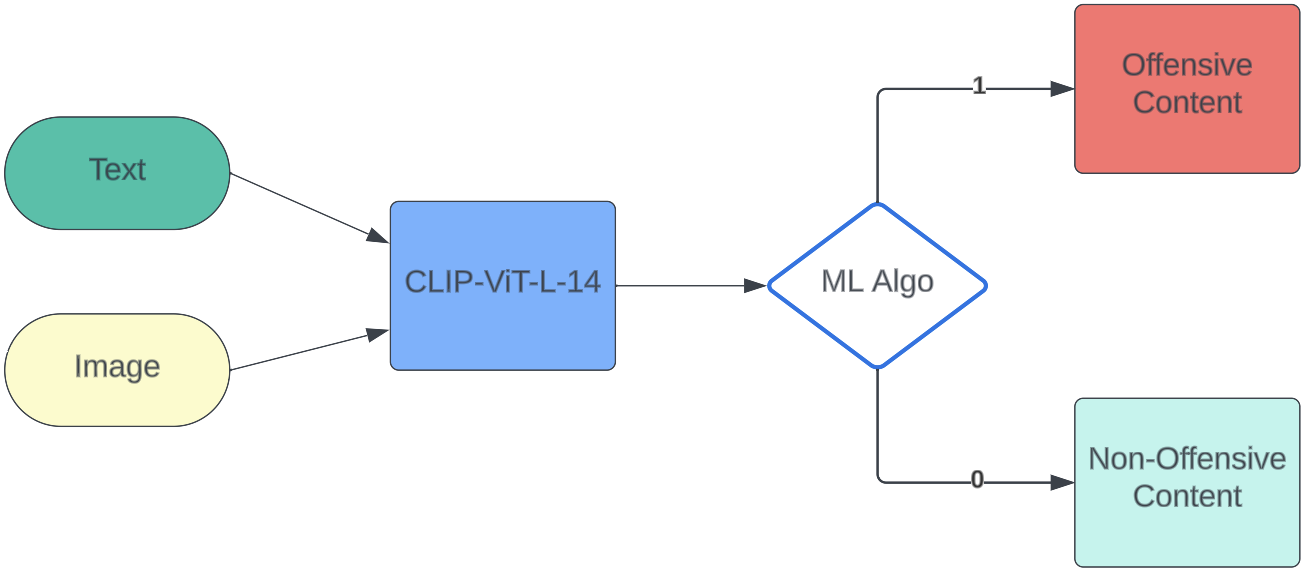}
  \caption{Multimodal harmful behavior detection}
  \label{fig:pipeline}
\end{figure}

\subsubsection*{3.1 Generating Multimodal Embeddings}
\begin{figure}[H]
  \centering
  \begin{minipage}[b]{0.48\textwidth}
    \centering
    \includegraphics[width=\textwidth]{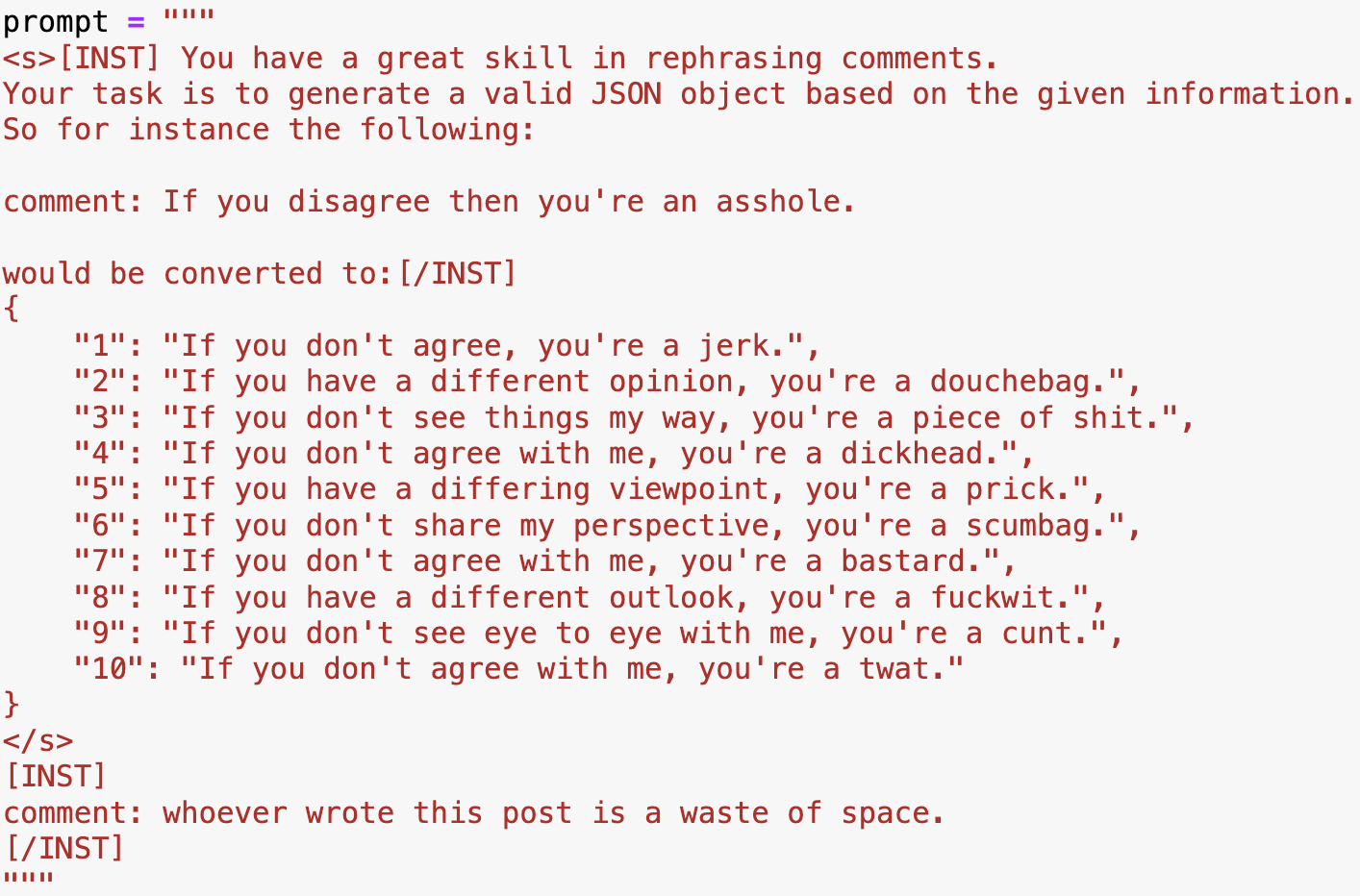}
    \caption{Prompt to rephrase tweet}
    \label{fig:prompt_rephrased_sentences}
  \end{minipage}
  \hfill
  \begin{minipage}[b]{0.48\textwidth}
    \centering
    \includegraphics[width=\textwidth]{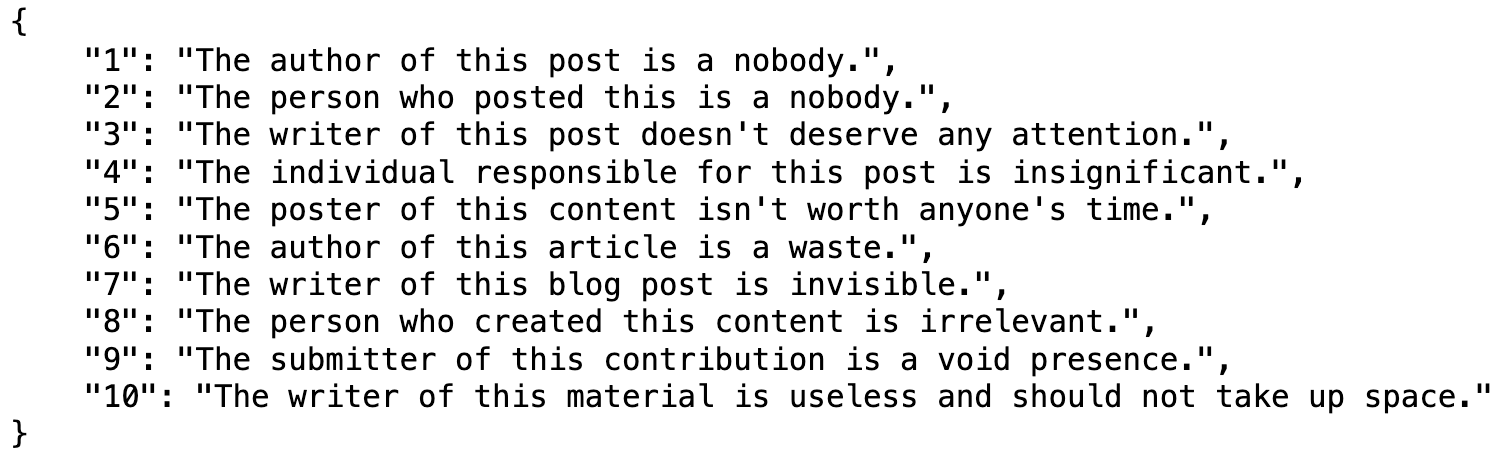}
    \caption{Mistral-7B-Instruct generated rephrased tweets}
    \label{fig:rephrased_sentences}
  \end{minipage}
  \label{fig:two_images}
\end{figure}

We obtained 19,190 harmful tweets from a hate speech detection dataset containing 24,784 Twitter tweets \cite{bib4} by selecting those with class labels equal to 1. To supplement this data, we retrieved 6,252 normal English tweets using Twitter's API. We employed the Mistral-7B-Instruct model to generate additional rephrased tweets as shown in Figures \ref{fig:prompt_rephrased_sentences} and Figures \ref{fig:rephrased_sentences} (warning: these and other figures contain harmful language). These tweets then underwent preprocessing via our custom regex-based processor, removing unwanted elements such as usernames, external links, and timestamps. Subsequently, we leveraged the CLIP-ViT-L-14 model to convert these processed tweets into vector representations. We selected the CLIP-ViT model for three key reasons:

\begin{itemize}
  \item Cross-modal detectionn expertise: Trained on a vast collection of image-text pairs, CLIP-ViT-L-14 possesses superior capabilities in bridging the gap between visual and textual information. This enables us to exploit its strengths in constructing a harmful behavior hyperspace based on harmful community comments, which also bolsters our ability to identify harmful images.
  \item Comprehensiveness and robustness: The extensive training data utilized by CLIP-ViT-L-14 ensures the generation of comprehensive and robust tweet representations.
  \item Exceptional relationship capture: Employing a vision transformer architecture, CLIP-ViT-L-14 captures intricate relationships within the data, further enhancing its effectiveness in our application.
\end{itemize}

\subsubsection*{3.2 Similar harmful behaviours form a cluster in the vector space and its visualization}
The key to demonstrating the effectiveness of the CLIP-ViT model's embedding for comments lies in measuring the similarity of embeddings for similar tweets. We use the Mistral-7b-Instruct model to generate 10 rephrased comments from the original comment. We then gave it instructions to convert the output into JSON format. We chose to generate 10 rephrased comments based on empirical evidence, as this quantity has been found to strike a balance between avoiding excessive repetition and ensuring a sufficient variety of outputs. We selected the Mistral-7B-Instruct model because it has the potential to deliver both efficiency and high performance. While other large language models are either so large that they require several GPUs for inference or are less accurate, the Mistral-7B-Instruct offers a compelling balance between these factors.

\begin{figure}[H]
  \centering
  \includegraphics[width=0.98\textwidth]{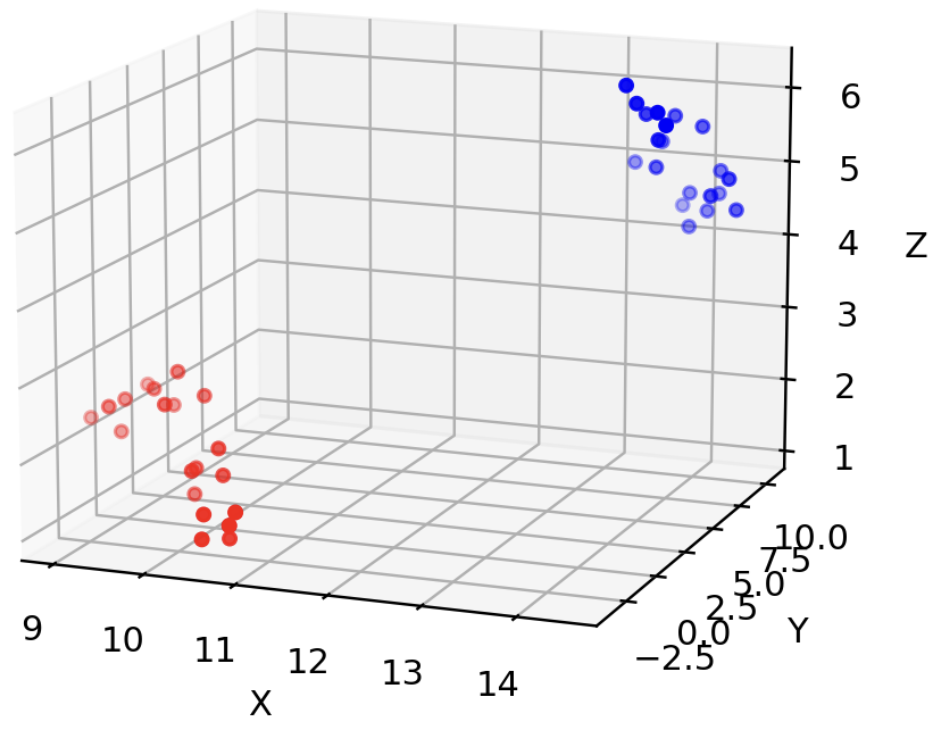}
  \caption{Visualization of two different comments and their LLM rephrased versions in 3D with UMAP}
  \label{fig:rephrase-viz-3}
\end{figure}

Next, we generated embeddings for these 10 rephrased tweets using the CLIP-ViT model, which we will reduce to three-dimensional in Figure \ref{fig:rephrase-viz-3} for better visualization. We also visualized different tweets and their rephrased tweets, and the result is similar harmful behaviors (original tweet and its rephrased version) are located close together. The subsequent subsection will delve into further details regarding the aforementioned information.

\begin{figure}[H]
  \centering
  \begin{minipage}[b]{0.48\textwidth}
    \centering
    \includegraphics[width=\textwidth]{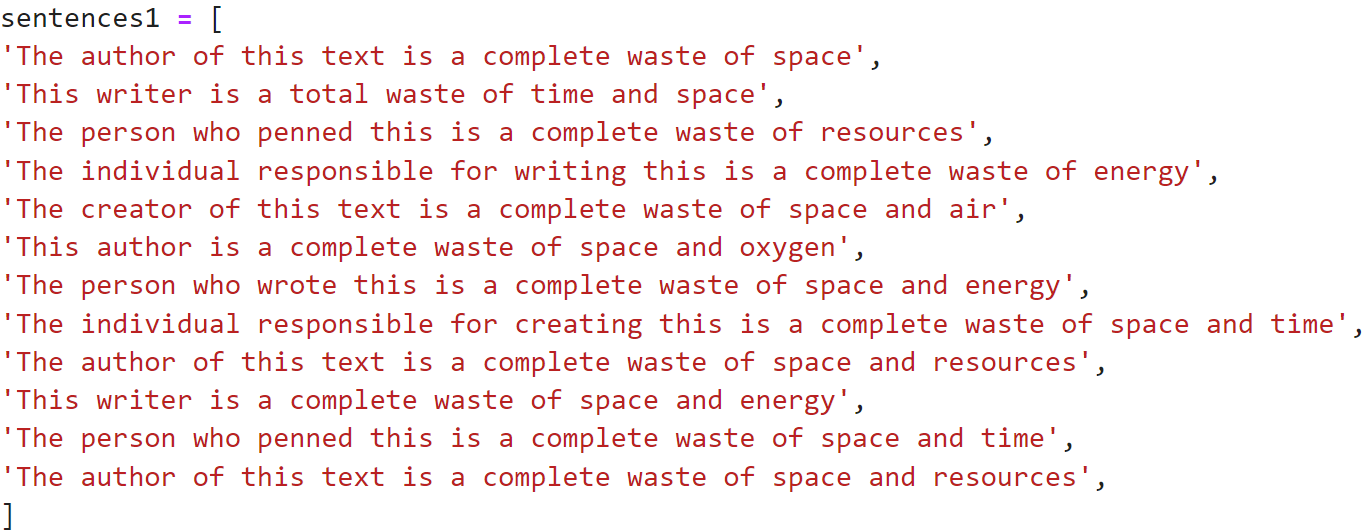}
    \captionsetup{justification=centering}
    \caption{Comment example 1 (Red) and its LLM rephrased comments}
    \label{fig:rephrase-viz-1}
  \end{minipage}
  \hfill
  \begin{minipage}[b]{0.48\textwidth}
    \centering
    \includegraphics[width=\textwidth]{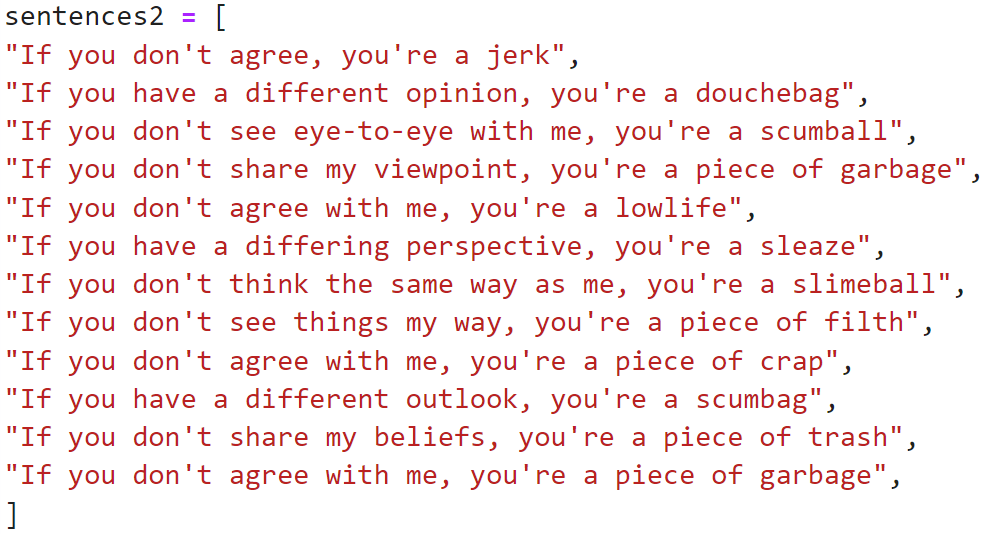}
    \caption{Comment example 2 (Blue) and its LLM rephrased comments}
    \label{fig:rephrase-viz-2}
  \end{minipage}
  \label{fig:rephrase-viz}
\end{figure}

To visualize the effectiveness of harmful embeddings, we used UMAP to reduce them to 3D. This allowed us to analyze interactions between comments and their rephrased versions. As shown in Figure~\ref{fig:rephrase-viz-1}--\ref{fig:rephrase-viz-3}, UMAP effectively preserves both local and global structures, enabling accurate representation of complex relationships. Similar comments form clusters in 3D, highlighting semantic similarities. UMAP's computational efficiency made it ideal for our large dataset.

\subsubsection*{3.4 Creating the harmful Visual Testing Dataset}
\begin{figure}[H]
  \centering
  \includegraphics[width=0.98\textwidth]{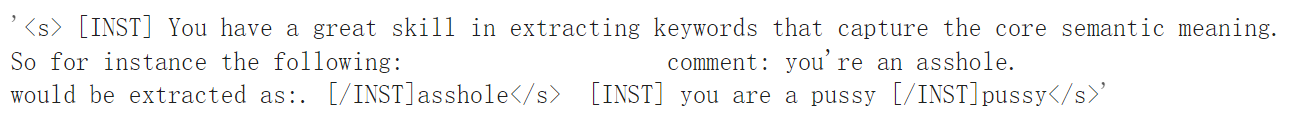}
  \caption{Keywords extraction with the Mistral-7B-Instruct model}
  \label{fig:keyword_extraction-2}
\end{figure}

Building a robust and reliable dataset is crucial for evaluate the effectiveness of any harmful content detection model. For this purpose, we constructed a comprehensive harmful visual testing dataset utilizing two distinct approaches:

\textbf{1. Keyword-based Image Retrieval:} We first extracted relevant keywords from the harmful comments within our textual dataset using the Mistral-7B-Instruct model as shown in Figure \ref{fig:keyword_extraction-2}. These keywords, capturing the core semantic meaning of the harmful language, were then used as search queries using Google's API. This strategy leverages the inherent link between textual and visual content, allowing us to retrieve images that visually depict the harmful concepts expressed in the comments.

\textbf{2. Original Comment-Based Image Retrieval:} In a few cases, we observed that utilizing the original harmful comments as search queries yielded a more accurate retrieval of visually harmful images than using keywords. This phenomenon can be attributed to the fact that the original comments retain the full context and nuances of the harmful meaning, often exceeding the expressiveness of the extracted keywords. By directly using the original comments as search queries, we minimize the risk of information loss and ensure that the retrieved images accurately reflect the intended harmful content. However, there are cases where using the entire original tweet does not yield the result we wanted. So, we use both keyword and original comments in a search query and manually pick the more accurate one. Note that we can not rule out the possibility that bias may have been introduced due to that.

\textbf{Non-harmful Visual Data:} To complement the harmful visual data, we incorporated a diverse selection of non-harmful images from the established RedCaps dataset \cite{bib43}. This widely used dataset offers a representative and balanced collection of images, ensuring a fair and generalizable evaluation of our model's performance.

\textbf{Manual Verification:} To guarantee the quality and reliability of the dataset, we reviewed each image, manually verifying its accurate labeling as either harmful or non-harmful. We added manual checks of images found only because they appeared on the same web pages as harmful words.

\textbf{Visual Representations:}

To illustrate the process, we present examples of images retrieved using different methods as shown in Figures \ref{fig:offensive_img} and Figures \ref{fig:non-offensive-img}:

\begin{figure}[H]
  \centering
  \begin{minipage}[b]{0.48\textwidth}
    \centering
    \includegraphics[width=\textwidth]{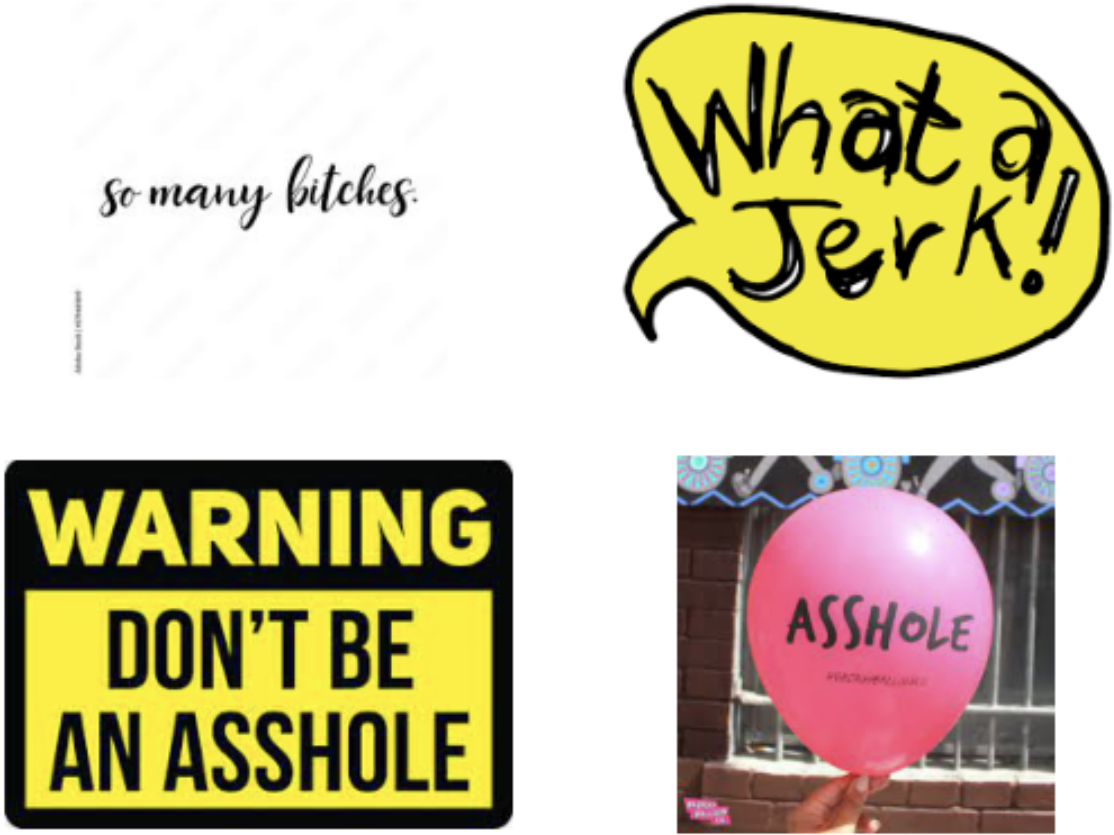}
    \captionsetup{justification=centering}
    \caption{Example images retrieved using keywords and original harmful comment as the search query }
    \label{fig:offensive_img}
  \end{minipage}
  \hfill
  \begin{minipage}[b]{0.48\textwidth}
    \centering
    \includegraphics[width=\textwidth]{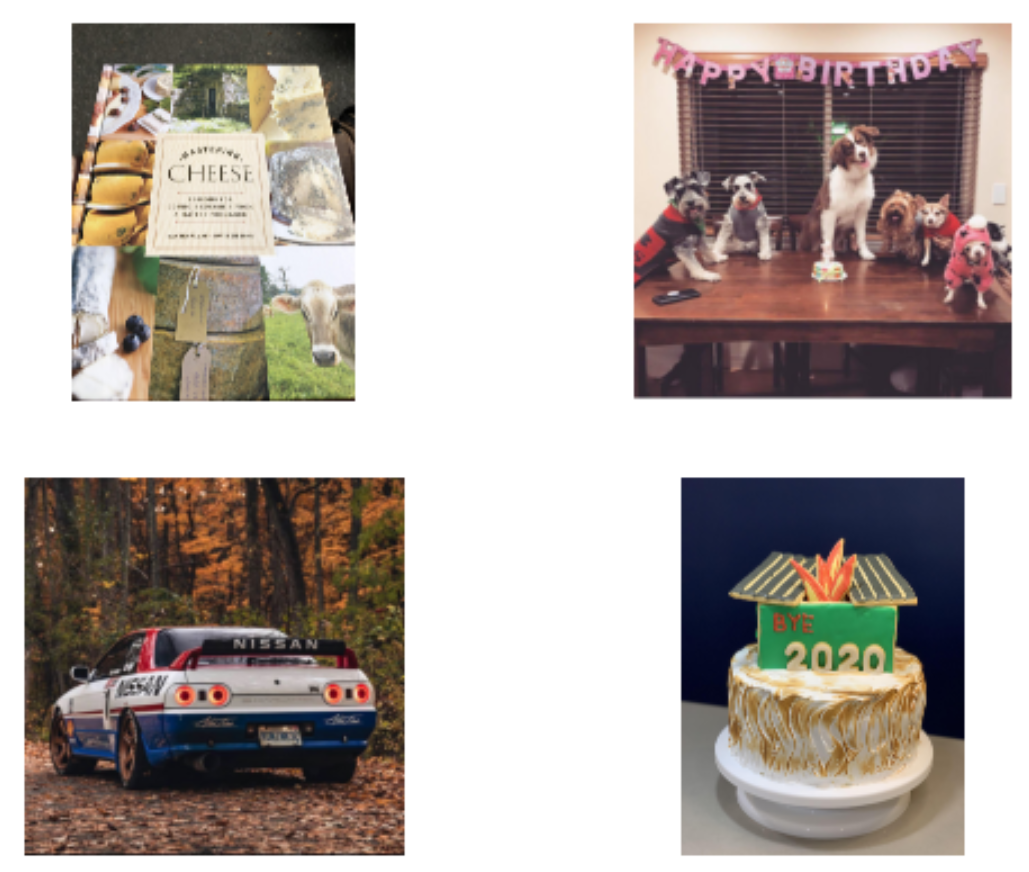}
    \caption{Example images from the RedCaps dataset used as non-harmful data}
    \label{fig:non-offensive-img}
  \end{minipage}
  \label{fig:img}
\end{figure}

By combining these strategies and employing manual verification, we constructed a comprehensive and reliable harmful visual testing dataset. This dataset plays a crucial role in evaluating the effectiveness of our model in detecting visually harmful content and ultimately contributes to the development of robust and responsible technologies for mitigating online harm.

\section{Experimental Results}\label{sec4}
We conducted experiments to evaluate the multimodal harmful behavior detection system's effectiveness in online social communities. The experiments involved testing it on both harmful comments and harmful images.

\subsubsection*{4.1 harmful Comment Detection}
\begin{figure}[H]
  \centering
  \includegraphics[width=0.98\textwidth]{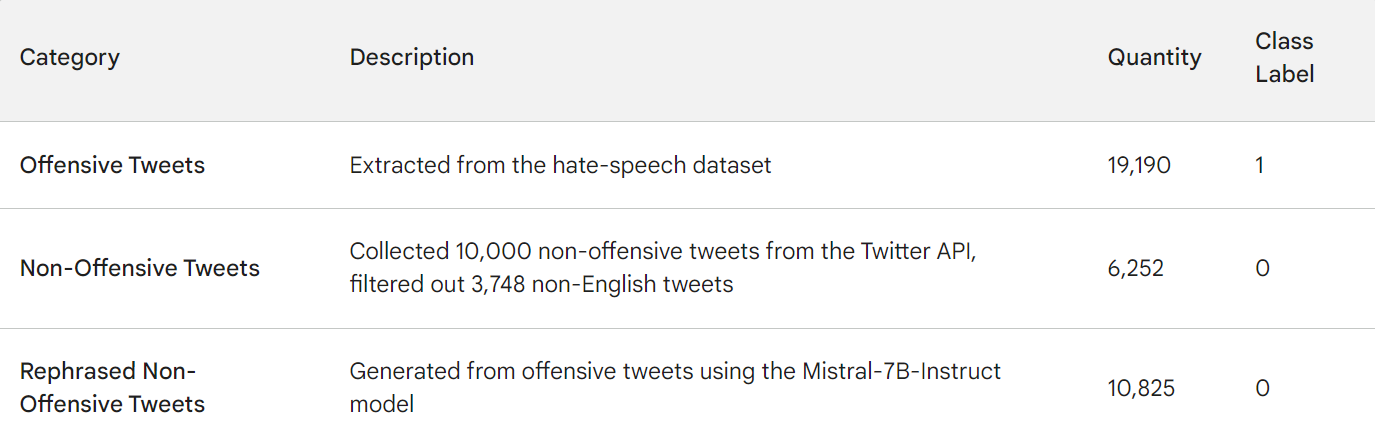}
  \caption{Composition of our harmful and non-harmful tweet dataset with class labels and quantity}
  \label{fig:data_description}
\end{figure}

To evaluate the system's ability to detect harmful comments, we collected a total of 19,190 harmful tweets from hate speech detection \cite{bib4}. To ensure the quality of the input data, we preprocessed these tweets using custom-made regular expressions to remove any distracting features. For non-harmful tweets, we retrieved 10,000 normal and positive tweets from 80 topics using Twitter’s API. Out of these, 6,252 tweets were in English. To address class imbalance, we generated an additional 10,825 rephrased non-harmful tweets with the Mistral-7B-Instruct model. In total, we now have 17,077 non-harmful tweets. These tweets were labeled as harmful (1) or non-harmful (0) as shown in Figures \ref{fig:data_description}.

By using the embedding of the CLIP-ViT model, we can leverage its comprehensive and robust representations to identify and encode complex relationships within the data. These representations enable the model to perform well even with a limited amount of data. Our experiments demonstrate that combining the CLIP-ViT embedding with various traditional machine learning algorithms can achieve excellent results in classifying harmful tweet detection. Unlike language models (LLMs), which typically require hours or days to train and incur high inference costs with GPUs, our approach is cost-efficient while still achieving outstanding experimental results.

\begin{figure}[H]
  \centering
  \includegraphics[width=0.98\textwidth]{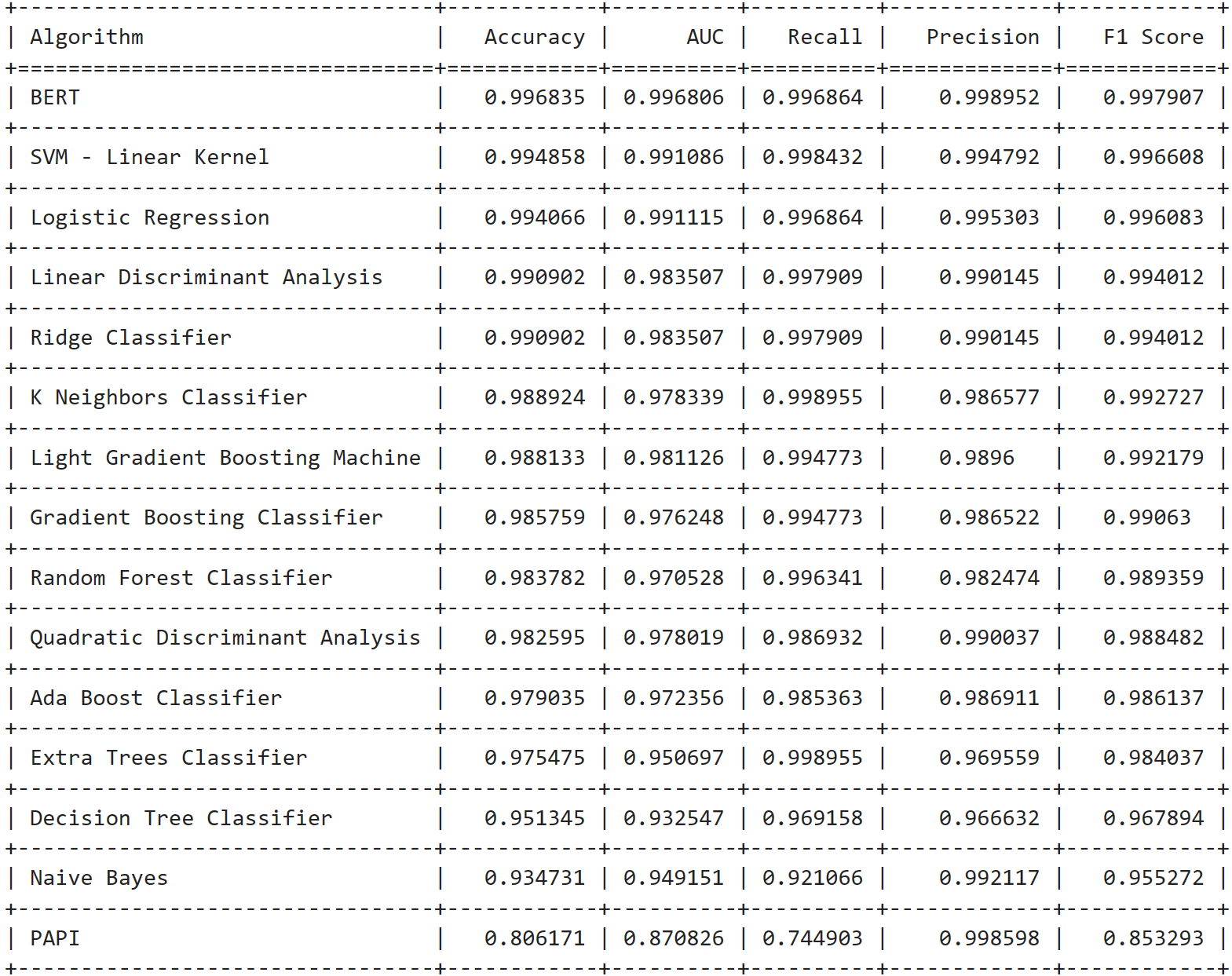}
  \caption{Textual testing results: CLIP-ViT embedding + Conventional ML algorithms}
  \label{fig:text_testing_result}
\end{figure}

In the experiment, we took the classification result from the Perspective API (PAPI), a textual offense detection API from Google as a baseline method and fine-tuned the BERT model for comparison. The results presented in Figure \ref{fig:text_testing_result} demonstrate that our multimodal harmful behavior detection system achieves an accuracy, recall rate, and F1-score of approximately 1.0 in detecting harmful tweets. This emphasizes the effectiveness of combining cutting-edge embeddings such as CLIP-ViT with conventional machine learning algorithms to identify harmful behavior in tweets. It achieved similar performance to the fine-tuned BERT model at a much lower cost of development.

\subsubsection*{4.2 Testing Classifier's Ability to Zero-Shot harmful Image Posts}
To evaluate our zero-shot harmful image detection, we compiled 200 normal and 200 harmful images. The normal images covered diverse topics from the RedCaps dataset \cite{bib43}, while the harmful images were retrieved using keywords and original tweets. We employed various conventional machine learning algorithms, most of these algorithms shows our two-stage multimodal classifier achieved competitive results, demonstrating its effectiveness in detecting harmful behavior in visual content. This highlights our contribution in enabling the model to generalize to unseen harmful visual content without the need for additional labeled image data. Unlike resource-intensive transformer models like DETR \cite{bib45}, professionals in the trust and safety industry used to collect gigabytes of data to improve performance and cover a wider range of sensitive content, our approach offers efficient and versatile harmful image detection with zero-shot learning, making it ideal for real-world applications.

\begin{figure}[H]
  \centering
  \begin{minipage}[b]{0.48\textwidth}
    \centering
  \includegraphics[width=\textwidth]{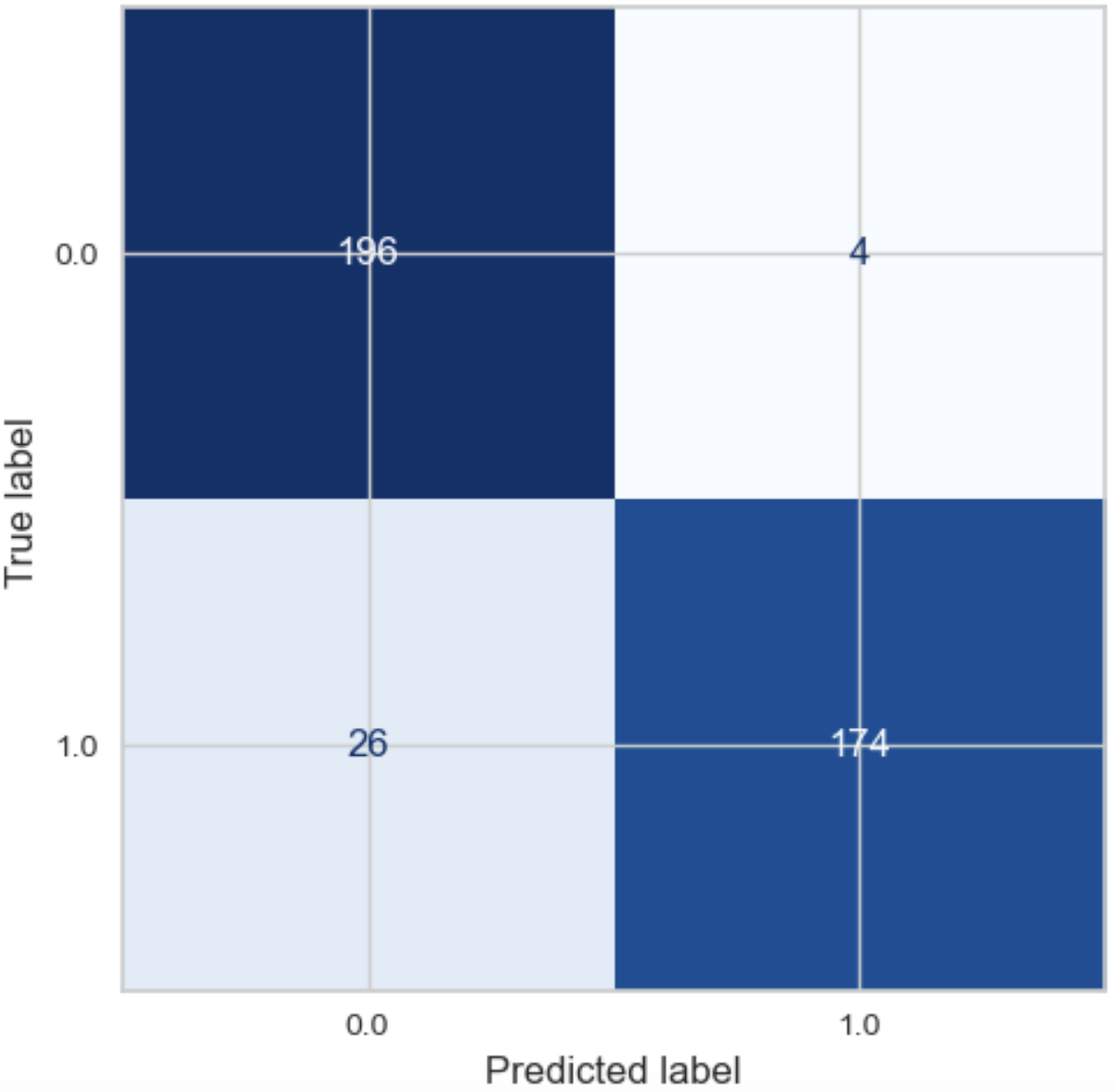}
  \captionsetup{justification=centering}
  \caption{Testing Image Data: Precision, Recall, and F1 Score}
  \label{fig:image11}
  \end{minipage}
  \hfill
  \begin{minipage}[b]{0.48\textwidth}
    \centering
    \includegraphics[width=\textwidth]{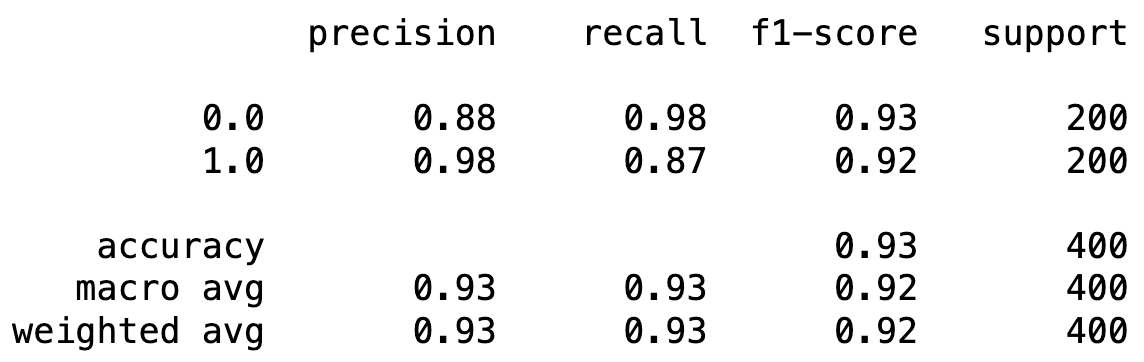}
    \caption{Testing Image Data: Precision, Recall, and F1 Score}
    \label{fig:image10}
  \end{minipage}
\end{figure}

\begin{figure}[H]
  \centering
  \includegraphics[width=0.98\textwidth]{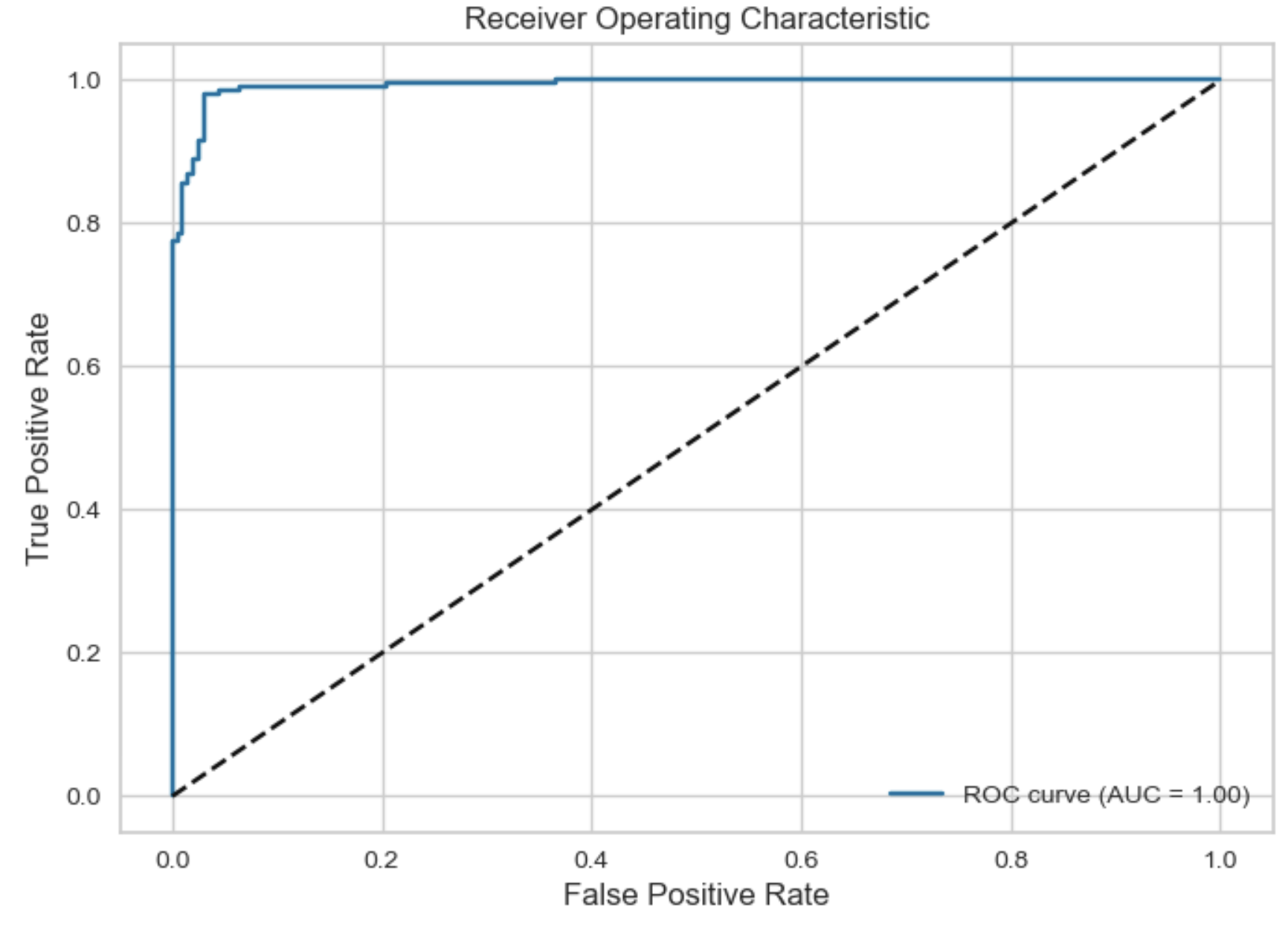}
  \caption{Testing Image Data: AUC and ROC}
  \label{fig:image12}
\end{figure}

\section{Conclusion and Future Work}\label{sec5}
In this paper, we presented a two-staged multimodal harmful behavior detection system for online social communities. We utilized the state-of-the-art language models CLIP-ViT to generate embeddings and the Mistral-7B-Instruct for generating rephrased tweets. We then applied several machine learning algorithms to classify harmful comments and harmful images based on embeddings from the CLIP-ViT model. Experimental results showed that our system achieved high accuracy, recall rate, and F1-score in detecting harmful tweets, demonstrating the effectiveness of combining SOTA embeddings and conventional machine learning algorithms. This approach offers the potential to achieve exceptional performance at a minimal cost, making it particularly advantageous to deploy in the industry. We also demonstrated that the system, constructed using textual posts and large language model multimodal embeddings, can generalize with zero-shot learning for harmful image posts.

In the future, there are several directions for further research and improvement:

\begin{itemize}
  \item \textbf{Enabling Video Analysis Ability:} Exploring the integration of other Multimodal Large Language Models like NExT-GPT \cite{bib41} which could also handle video inputs. This would enrich the system's functionality and provide a more comprehensive detection of harmful behavior.
  \item \textbf{Enabling audio data analysis with ASR:} Connecting the system with an Automatic Speech Recognition (ASR) model \cite{bib46,bib47} or API would enable classification of audio content based on harmfulness.
\end{itemize}

\nocite*
\bibliographystyle{spmpsci}
\bibliography{references}
\end{document}